\documentclass[sigconf]{acmart}

\usepackage{array}

\usepackage{listings}

\usepackage{xcolor}

\definecolor{codegreen}{rgb}{0,0.6,0}
\definecolor{codegray}{rgb}{0.5,0.5,0.5}
\definecolor{codepurple}{rgb}{0.58,0,0.82}
\definecolor{backcolour}{rgb}{0.95,0.95,0.92}

\lstdefinestyle{mystyle}{
  backgroundcolor=\color{backcolour}, commentstyle=\color{codegreen},
  keywordstyle=\color{magenta},
  numberstyle=\tiny\color{codegray},
  stringstyle=\color{codepurple},
  basicstyle=\ttfamily\footnotesize,
  breakatwhitespace=false,         
  breaklines=true,                 
  captionpos=b,                    
  keepspaces=true,                 
  numbers=left,                    
  numbersep=5pt,                  
  showspaces=false,                
  showstringspaces=false,
  showtabs=false,                  
  tabsize=2
}

\AtBeginDocument{%
  }

\setcopyright{acmlicensed}
\copyrightyear{2025}
\acmYear{2025}
\setcopyright{acmlicensed}\acmConference[SIGCSE TS 2025]{Proceedings of the 56th ACM Technical Symposium on Computer Science Education V. 1}{February 26-March 1, 2025}{Pittsburgh, PA, USA}
\acmBooktitle{Proceedings of the 56th ACM Technical Symposium on Computer Science Education V. 1 (SIGCSE TS 2025), February 26-March 1, 2025, Pittsburgh, PA, USA}
\acmDOI{10.1145/3641554.3701874}
\acmISBN{979-8-4007-0531-1/25/02}

\begin{document}

\title{Drafter: A Python Library for Full-Stack Web Development in CS1}

\author{Austin Cory Bart}
\email{acbart@udel.edu}
\affiliation{%
  \institution{University of Delaware}
  \city{Newark}
  \state{DE}
  \country{USA}
}

\author{Nazim Karaca}
\email{nazim@udel.edu}
\affiliation{%
  \institution{University of Delaware}
  \city{Newark}
  \state{DE}
  \country{USA}
}

\renewcommand{\shortauthors}{Bart and Karaca}

\begin{abstract}

Web applications are increasingly the main way to create user interfaces, and they are often the most common software beginners have encountered. However, web development relies on concepts like HTML, CSS, JavaScript, and backend technologies, which are challenges beyond the scope of an introductory course. Additionally, many features of modern web frameworks conflict with CS1 principles, such as avoiding global mutable state and promoting test-driven development. Consequently, web development is rarely integrated into CS1 courses, despite its motivational potential.

This paper introduces Drafter, a new open-source Python library for CS1, enabling students to develop full-stack web applications using pure functions with minimal boilerplate. Drafter has simple functions to generate HTML, eliminating the need to learn HTML or templating languages. The data model scales with students' mastery of types, from primitives to nested lists, Dataclasses, and dictionaries. Drafter supports unit testing of entire web applications and reinforces CS1 concepts like decomposition and reusability. Websites created with Drafter can be deployed through GitHub Pages, with plotting and image manipulation. Students receive extensive debug information, including unit test scaffolding, program state visualization, and enhanced error messages and tests.

This paper discusses the pedagogical features and design decisions behind Drafter, along with instructor reflections on its use. We believe Drafter can significantly enhance student engagement in CS1 courses and reinforce key learning objectives.

\end{abstract}




\maketitle

\section{Motivation}

Novice computer science students often have extensive experience with interactive web applications prior to their first computing course.
Despite this familiarity, teaching web development in CS1 courses presents significant challenges, and so is an uncommon context.
Web development requires HTML, CSS, JavaScript, and backend technologies -- topics that are often too complex for beginners.
Additionally, many web frameworks conflict with fundamental CS1 principles, such as promoting global mutable state.
These frameworks are typically heavy with boilerplate and have steep learning curves for their core concepts.

To address these challenges, we created Drafter~\footnote{\url{https://drafter-edu.github.io/}}, an open-source Python library designed for CS1 students to make web applications.
Drafter allows students to create ``full-stack'' web applications using pure functions with minimal boilerplate.
The library provides user-friendly functions for generating HTML and manages state, enabling students to focus on core programming concepts without the overhead of learning additional languages or frameworks.

\subsection{Contributions and Objectives}

Our primary design goals for Drafter were to:

\begin{enumerate}
    \item Empower novices to create rich graphical applications
    \item Discourage bad practices like global, mutable state
    \item Promote good practices like decomposition and unit testing
    \item Minimize learners' cognitive load, by requiring mastery of only a few key concepts
    \item Avoid the syntax of additional language (i.e., HTML, CSS)
\end{enumerate}

This paper makes a number of contributions, including:

\begin{enumerate}
    \item Identification of design issues for novice web frameworks
    \item Descriptions of the major features of Drafter with examples
    \item Explanations of how Drafter can be taught
    \item Reflection on our experiences using Drafter in a CS1
\end{enumerate}

Fundamentally, we believe that Drafter is a novel tool designed to bridge the gap between introductory programming education and modern web development practices, making it an effective and engaging learning experience for CS1 students.

\section{Prior and Related Work}

We reviewed the literature on both web development in computer science degrees generally, and also specifically in CS1 contexts.
We then also reviewed existing web frameworks for Python, hoping to find a suitable pre-existing solution to our design needs.
In both cases, we found existing solutions inadequate for addressing our specific educational goals.

\subsection{Research on Web Development in CS1}

Generally, Computer Science programs have had limited integration of web development.
For example, the ACM Curriculum 2023 identifies web development as a ``Knowledge Area Core'' (i.e., optional)~\cite{acmCurriculum2023} instead of a ``Computer Science Core'', suggesting it as an optional context.
Connolly, Alston, and others~\cite{connolly2019facing,miller2015introduction,alston2015uncovering} have been vocal advocates for Computer Science Education to integrate web development more deeply into the curriculum, arguing as recently as 2019 that its role is paramount in the curriculum as users increasingly on web applications to interact with their computer (compared to desktop or command line technologies).
These papers do acknowledge a number of (reasonable) issues that have impeded the spread of web development as an introductory context:

\begin{itemize}
    \item The rapid evolution of Web technology makes it almost impossible to keep curricular materials relevant and up-to-date, with APIs and standards changing even during a semester.
    \item The relative novelty of web programming paradigms compared to conventional paradigms like Object-Oriented Programming, data structures, and other historically core topics; we know less about how to teach this subject effectively compared to the classics.
    \item Limited consensus on what exactly constitutes ``web development'', and the huge breadth of possible topics to include (backend frameworks, frontend frameworks, network communication protocols, database technologies, etc.).
    \item The inherently complex nature of these technologies, which require complex interactions and deployment workflows for anything but the most trivial websites.
\end{itemize}

It is not surprising then that the limited prior research we have found on introductory web development tends to be disparate and scattered.
Most web development education occurs in advanced courses~\cite{barzilai2023using,wang2013web,dugan2013single}.
We have not found any literature on a Python-based CS1 that incorporates web development, although we did find experiences in other languages.
For example, Schaub~\cite{schaub2009teaching} developed a Java CS1 curriculum that teaches back-end web development.
Hamid~\cite{hamid2012automated} developed a Scheme CS1 that works somewhat similar to Drafter, and mirrors many of our functional approaches.
There have also been some attempts to teach web development in follow-up courses, such as Steppe's CS1.5 course that served as a follow-up to their CS1, taught in JavaScript.
However, none of these courses seemed to have taught web development in a Python CS1, likely due to the complexity of Python web development frameworks, as we discuss in the next subsection.

\subsection{Existing Python Web Development Libraries}

Before developing Drafter, we reviewed various modern web development libraries available on PyPI~\footnote{\url{https://pypi.org/}}. However, none met our design goals or were built with true novices in mind. The main issues included:

\begin{description}
    \item[Global State] Most web applications rely on global mutable state, often mediated through special objects that rely on dynamic contextualization (e.g., \texttt{Flask}'s \texttt{g} object, \texttt{Bottle}'s \texttt{request}, \texttt{Pyneone}'s \texttt{pc.State}).
    \item[Requires HTML/CSS] Most frameworks assume that the user is comfortable writing raw HTML and CSS. For students, this increases the cognitive load since they are already focused on learning Python. Although these markup languages are comparatively simple to programming languages, they still incur overhead that distracts from the core learning objectives.
    \item[Templating Languages] Some frameworks rely heavily on templating languages (e.g., \texttt{Web2Py}, \texttt{Jinja2} for \texttt{Flask}), which extend HTML with programming language features. These extensions are powerful, but increase the learning curve.
    \item[Backend Only] On the other end of the spectrum, some libraries only provide mechanisms for generating \texttt{JSON} responses and are not focused on developing user interfaces (e.g., \texttt{FastAPI}). We still wanted novices to create real graphical user interfaces, so these libraries were inadequate.
    \item[Class-based Routing] Some frameworks (e.g., \texttt{CherryPy}) primarily focus on or only provide class-based routing, which increases the prior knowledge required to begin development. Worse, the structure of those classes can be at odds with the patterns of classes taught in introductory Object-Oriented courses, or rely on even more advanced structural patterns like function closures (e.g., \texttt{FastHTML}).
    \item[Boilerplate Code] Although some libraries require only decorators and a \texttt{run} call to make things work. However, other frameworks like \texttt{Pyramid} require more substantial setup (including use of more advanced control structures like \texttt{with}).
    \item[Complicated Project Setup] Web applications are often sprawling programs with multiple files, components, and layers. Some web frameworks (e.g., \texttt{Django}) help organize these files by providing specialized command line tools to initialize the repository. This requires novices to learn how to use command line tools and navigate their filesystem, even for a beginning application. Other tools like \texttt{Anvil} provide a specialized IDE with its own learning curve.
    \item[Asynchronous] Many modern frameworks (e.g., \texttt{Starlette}) are built for performance, which can mean building the application to run asynchronously. This means that functions must be annotated with the \texttt{async} keyword, necessitating either more hand-waving or more explanations.
\end{description}

Many of these topics are excellent learning opportunities that could be presented to novices with appropriate scaffolds.
However, we believed the sum of them to be overwhelming, and motivated our decision to develop Drafter.

\section{Approach}

Fundamentally, Drafter provides a set of functions for creating simple web applications described with more detail in \ref{lbl:api}.
Then, in \ref{lbl:features} we describe how these primitives are enhanced further for the pedagogical context.

\begin{figure}
    \centering
\begin{lstlisting}[language=Python]
from drafter import Button, TextBox, Page
from drafter import route, start_server, dataclass

@dataclass
class State:
    first_number: str
    second_number: str
    result: str

@route
def index(state: State) -> Page:
    content = [
        "Give two numbers:",
        TextBox("first", state.first_number),
        TextBox("second", state.second_number),
        Button("Add", add)
    ]
    if state.result:
        content.append("The result is: " + state.result)
    return Page(state, content)

@route
def add(state: State, first: str, second: str) -> Page:
    state.first_number = first
    state.second_number = second
    if first.isdigit() and second.isdigit():
        state.result = str(int(first) + int(second))
    else:
        state.result = "Invalid numbers!"
    return index(state)

start_server(State("", "", ""))
\end{lstlisting}

    \includegraphics[width=1\linewidth]{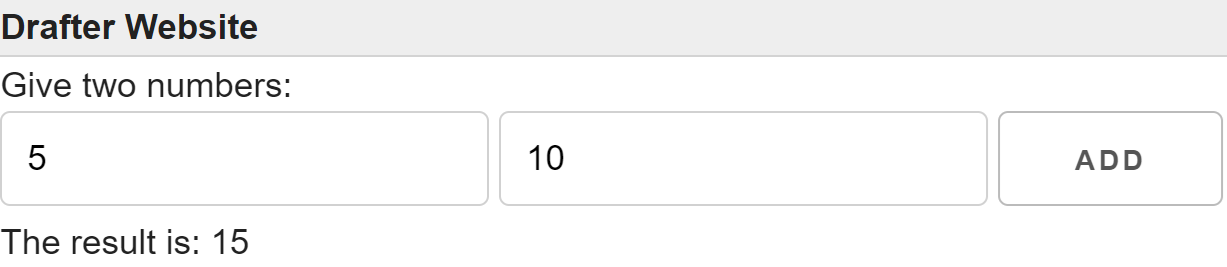}
    \caption{Simple Calculator Example}
    \label{fig:calculator}
\end{figure}

\subsection{Drafter's Core API}

\label{lbl:api}

The three major built-in functions required for Drafter are \texttt{start\_server}, \texttt{route} (a decorator), and \texttt{Page} (a dataclass).
Users are typically expected to define their own \texttt{State} dataclass.
They will also take advantage of a large collection of built-in component \texttt{dataclasses} that correspond with HTML elements and CSS styling operations.

\subsubsection{The \texttt{start\_server} Function}

The main function for any Drafter application is \texttt{start\_server}, which actually launches the server. In the desktop version of the application, this creates a \texttt{Bottle} server (which can actually be reconfigured via keyword parameters to the \texttt{start\_server} function, e.g., to change the port). This call must be made as the last step of the program (even after all the unit tests), since the server loop blocks execution.

\subsubsection{The \texttt{route} Function}

The main mechanism for creating web pages in Drafter is to define functions that are annotated with the \texttt{@route} decorator. The function takes a copy of the current \texttt{State} object as the first argument, and must return a \texttt{Page} object.
Since routes are just simple functions, route functions can be reused in other routes (as shown in line 31 of Figure \ref{fig:calculator}, where the rendering logic of the \texttt{index} route is reused in the \texttt{add} route).

Route functions are intended to be \textit{pure} (in the functional programming sense), without additional side-effects or global state (mutation of the copied \texttt{state} parameter is allowed).
By using pure functions, Drafter helps students develop a better understanding of functional programming principles such as immutability and referential transparency. This not only improves code quality but also makes it easier for students to reason about and test their programs.

\subsubsection{The \texttt{State} Class}

Most web servers maintain some state between page accesses. In the Calculator example in Figure \ref{fig:calculator}, the state is three numbers. For an adventure game, it might be the player's health and inventory. For an e-commerce application, the state could be a shopping cart. Therefore, Drafter manages a \texttt{State} object (defined by the programmer) between requests.

The \texttt{State} class is actually not an intrinsic part of the Drafter library, but is a convention whereby the developer defines a \texttt{State} class and then provides it as the first parameter for their route functions and the first argument to their \texttt{Page} instances.
Technically, you do not have to create a \texttt{State} class; you are free to use other types such as lists, dictionaries, and primitives (e.g., strings, integers, booleans). You are even free to have no state at all (as in figure \ref{fig:stateless}).
This allows Drafter to be used prior to teaching any kind of Object-Oriented Programming, or even without any Object-Oriented Programming at all!

\begin{figure}
    \centering
\begin{lstlisting}[language=Python]
from drafter import *

@route
def index() -> Page:
    return Page(["Hello world!"])

start_server()
\end{lstlisting}

    \caption{A Minimal, Stateless Drafter Website}
    \label{fig:stateless}
\end{figure}

\subsubsection{The \texttt{Page} Class}

The \texttt{Page} constructor takes two parameters, the new \texttt{state} and the \texttt{content} to render.
If the \texttt{state} is unchanged, it can be passed in unmodified. To change the state, you can either use an immutable approach of constructing a new \texttt{State} object (passing in the original fields), or use attribute assignment to modify the parameter (line 25 of \ref{fig:calculator}).
The \texttt{content} is a list of strings and component instances that will be the HTML of the page.
The strings can contain arbitrary HTML (and therefore CSS and JS); however, the expected approach is to use components as shown in lines 15 through 17 of Figure \ref{fig:calculator}.

\subsubsection{The Component Classes}

Table \ref{tbl:components} lists the components and corresponding HTML element.
Optional arguments have a question mark, and variadic arguments begin with ellipses.

\begin{table}[h!]
\centering
\begin{tabular}{>{\ttfamily}l >{\ttfamily}l}
\textbf{Function} & \textbf{HTML Tag} \\
\hline
Argument(name, value) & input \\
BulletedList(items) & ul \\
Button(text, url, arguments?) & button \\
CheckBox(name, default\_value?) & input \\
Column(...items) & div \\
FileUpload(name) & input \\
Header(text, level?) & h1--h6 \\
HorizontalRule() & hr \\
Image(url, width?, height?) & img \\
LineBreak() & br \\
Link(text, url) & a \\
MatPlotLibPlot() & img \\
NumberedList(items) & ol \\
Row(...items) & div \\
SelectBox(name, options, default\_value?) & select \\
Span(...items) & span \\
Table(data) & table \\
TextArea(name, default\_value?) & textarea \\
TextBox(name, default\_value?) & input \\
\end{tabular}
\caption{The Major HTML Components Available in Drafter}
\label{tbl:components}
\end{table}

Some components merit further explanation:

\begin{itemize}
    \item With \texttt{Link} and \texttt{Button} components, routes can be linked to other routes either by their function object (as shown on line 17 of Figure \ref{fig:calculator}) or the string name of the function.
    \item The \texttt{Image} component can take either the name of a local image file or an absolute URL.
    \item The \texttt{Button} component can also take a list of \texttt{Argument} components, which allows you to attach additional data to send when the button is pressed.
    \item On its own, the \texttt{Argument} component can be used to embed hidden fields into the page (usually controlled, e.g., by conditional logic).
    \item The \texttt{Table} class can take a list of dataclasses or a single dataclass and visualize the data in an HTML table.
    \item The \texttt{MatPlotLibPlot} serves as a drop-in replacement for \texttt{plt.show()} so that students can generate histograms, line plots, and bar graphs.
\end{itemize}

\subsubsection{The Styling Functions}

A final set of functions in the Drafter API is a large collection of functions built to manipulate the style and attributes of components.
There are functions for changing color (e.g., \texttt{change\_color}), font size (e.g., \texttt{change\_text\_size}), margin (e.g., \texttt{change\_margin}), padding (e.g., \texttt{change\_padding}), and any other CSS characteristics that novices might reasonably need to modify.
All of these functions take in a component and return the same component modified, allowing you to chain the style transformations.
Additionally, every component accepts optional keyword arguments (prefixed with \texttt{style\_}, e.g., \texttt{style\_font\_size}).

\subsection{Major Drafter Features}

\label{lbl:features}

Drafter provides several key features designed to simplify web development for CS1 students and reinforce core Computer Science, Software Engineering, and programming concepts.
During development, Drafter websites launch as a web page with additional debug information at the bottom.
Since web applications are naturally browser-based, the tool leverages the existing interface to provide additional information and utilities.
Internally, Drafter has other features that provide additional pedagogical support to students.
Below, we elaborate on these features and their benefits.

\subsubsection{Graphical Test Diffs}

Drafter routes can be tested with any Python testing libraries (e.g., \texttt{unittest}, PyTest), but also has a testing function (\texttt{assert\_equal}).
Figure \ref{fig:tests} demonstrates how you can write concrete tests for each route function, where students simply specify the next \texttt{State} and generated \texttt{Page} content list.
In addition to outputting on the command line, this function will embed its results in the debug area of the browser window.
This allows us to provide rich visual diffing (highlighting extraneous data in green, unnecessary data in red, and modified data in purple).

\begin{figure}
    \centering
\begin{lstlisting}[language=Python]
# ... Assume that the code from Figure 1 is reproduced...
from drafter import assert_equal

assert_equal(index(State("1", "7", "")),
             Page(State("1", "7", ""), [
              "Give two numbers:", TextBox("first", "1"), 
              TextBox("second", "7"), Button("Add", add)
             ]))
assert_equal(add(State("", "", ""), "4", "2"),
             Page(State("4", "2", "6"), [
              "Give two numbers:", TextBox("first", "4"),
              TextBox("second", "2"), Button("Add", add),
              "The result is: 6"
             ]))

start_server(State("", "", ""))
\end{lstlisting}

    \caption{Unit Tests for the Calculator Example}
    \label{fig:tests}
\end{figure}

\subsubsection{Scaffolded Unit Test Construction}

Drafter tracks navigation, and then automatically lists route accesses at the bottom of the page as a suite of ready-to-go unit tests.
This is helpful for regression tests when students are happy with the design of their site, but also helpful for quickly scaffolding lengthier tests.
Novices often struggle to write tests~\cite{bai2021testing}, partially due to the inherent difficulty of testing (which requires critically confronting your own code) but also because of students' negative perceptions about testing (that it is difficult or unhelpful~\cite{janzen2007perceptions}).
There is research to suggest that at least requiring students to write tests later (as opposed to earlier) may have some motivational benefits and help them along the path to adopting test-driven development~\cite{janzen2008test}.
We believe that making it easier to more interactively test their application will improve student attitudes towards testing, especially when combined with intentional classroom activities that highlight the value of testing.

\subsubsection{Program Visualization}

\begin{figure}
    \includegraphics[width=1\linewidth]{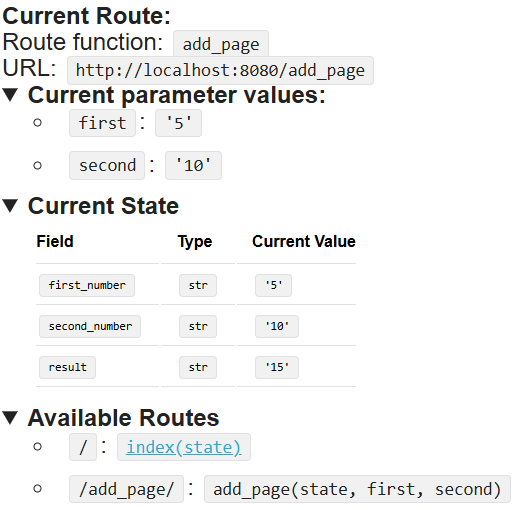}
    \caption{Subset of Debug Information for Calculator}
    \label{fig:debug}
\end{figure}

In the debug area, Drafter provides a tabular view of the current application state, including the types and values of all of the \texttt{State}'s fields.
The current endpoint and its parameters are also listed to show how the current page was reached.
Additionally, all endpoints available in the application are listed along with their associated route function.
An example of some of this data is shown in Figure \ref{fig:debug}.
Finally, a list of all accessed routes (along with their arguments and page content) is shown as a record of the pages accessed; users can click on any of the history to jump back to that point in time.

\subsubsection{Serializable State}

Drafter has sophisticated mechanisms for storing state across pages without relying on binary serialization (i.e., it does not need to \texttt{pickle}).
These features even work for complex objects like Dataclasses, using Python's new static typing features.
Drafter inspects the names and types of route parameters in order to align data from forms with the next endpoint.
Drafter will also intuitively coerce types where meaningful, e.g., converting checkboxes to boolean parameters and number boxes to integers.

\subsubsection{Internal Validation and Enhanced Errors}

Drafter includes built-in validation mechanisms to ensure that data passed between components and routes is of the correct type and shape (without extraneous, missing, or mistyped parameters and fields). This internal validation helps catch errors early and provides informative feedback to students.
For example, if \texttt{Page} is constructed with an improper type of value in its content list (e.g., a dictionary), then Drafter will generate a clear error message indicating this mismatch (\textit{``The content of a page must be a list of strings or components. Found a \texttt{dictionary} at index 5 instead}). Or if an endpoint expected a certain number of parameters, but the student forgot to provide one via a \texttt{TextBox} or \texttt{Argument} field (\textit{The \texttt{save\_changes} function expected 3 parameters, but 2 were provided. Expected: \texttt{state}, \texttt{new\_name}. But did not receive: \texttt{password}}). This feature reduces the likelihood of runtime errors and helps students understand the importance of type safety and data validation in their applications.

\subsubsection{Client-side Execution}

Web applications typically require a backend server to handle requests, manage state, and serve content.
In addition to deploying as a regular backend application, Drafter sites can also run entirely in the user's browser via a client-side execution library (built in Skulpt~\footnote{\url{https://github.com/skulpt/skulpt/}}), meaning there is no need for a backend server.
This approach allows students to deploy using static hosting services like GitHub Pages~\footnote{\url{https://pages.github.com/}}, providing students with a real-world context for their projects.

\subsection{Documentation and Tutorials}

A library, no matter how well-designed, is not sufficient on its own as a pedagogical resource, but must be paired with educational materials that provide students with opportunities for participation and feedback.
To that end, we have created a number of lessons and assignments for teaching students how to use Drafter:

\begin{itemize}
    \item Quick Start Guide: A visual tutorial that steps students through creating a basic web application with forms, multiple endpoints, and unit tests.
    \item Workbook: A set of four skeletons for web applications along with tutorial text that explains how to complete the tests, with unit tests provided to automate the grading. Students develop a Cookie Clicker-like game~\footnote{\url{https://en.wikipedia.org/wiki/Cookie_Clicker}}, a Bank Account simulator, a simple choose-your-own adventure Game, and an item shop for an RPG fantasy game.
    \item Longer Examples: A collection of a half-dozen more complicated worked examples including a calculator, login form, and dog-tracking system.
    \item Deployment Guide: A visual walkthrough of all the steps required to deploy an existing Drafter site on GitHub pages, 
\end{itemize}

These resources are all available on the Drafter website~\footnote{\url{https://drafter-edu.github.io/}}, along with our official API documentation.
In fact, Drafter provides two versions of its documentation: a detailed version for developers, and a more minimal version tuned for novices using simpler vocabulary and hiding unnecessary details.

\section{Evidence}

In this section, we describe our experiences using Drafter in an actual CS1 course.
We reflect on some of the successes and issues we had, and what we plan to do differently going forward.

\subsection{Context}

Drafter was used in two introductory courses in the 2023-2024 academic school year at an R1 university on the eastern United States coast across two semesters.
One of the paper's authors was the instructor of the course, and the other was heavily involved as an course designer.
The course is a Python CS1 that uses the Bakery curriculum, which has more information available in \cite{bart2024cs1}.
This curriculum emphasizes core principles of Computer Science, fundamentals of programming, and the basics of Software Engineering.
Functions are taught early, but classes are only taught as Data Classes (without methods or inheritance).
Significant weight is given to testing, code quality, and planning.
Each semester had hundreds of students using Drafter, although not every introductory section incorporated Drafter (instead using other contexts).
One of the CS1 courses were for Computer Science majors, and the other course a CS1 was for Engineers.

The biggest integration point for Drafter in these courses was the final project, which was a four week endeavor with multiple milestones and deliverables before the final web application.
The project itself was very open-ended, with students free to make any web application that they wanted, as long as the code met certain technical requirements.
These requirements included but were not limited to: having at least 7 routes, 5 distinct pages, a \texttt{State} with at least 4 fields (one of which was a list) that were meaningfully modified, at least 4 input fields (e.g., \texttt{SelectBox}, \texttt{TextBox}).
Additionally, students were required to provide tests, static types, and avoid the use of any global variables.

To prepare students for the final project, we had two lab assignments based heavily on the Quick Start Guide (week 7) and the Drafter Workbook (week 10).
Then, the last five weeks of the semester were dedicated to the final project itself.
First, students were required to develop a graphical representation of the plan for their website, identifying the key routes and state.
Then, over the next few weeks, they were expected to complete two milestones that would demonstrate incremental development -- an autograding script would confirm that they had made progress towards the requirements (e.g., the first milestone required at least one-third of the routes, and the second milestone required at least two-thirds of the routes).
The final project submission was expected to be deployed through GitHub Pages using the previously mentioned guide, and was graded with a rubric.

The instructors dedicated four lecture sessions to the project, providing useful spaces for troubleshooting.
This also gave in-class time for students to discuss and share with their peers, which helped build community and a sense of belonging, as well as normalizing the struggles of bug-hunting and -fixing.
Otherwise, students were expected to work on the final project outside of class.

\subsection{Instructor Experiences}

In this initial deployment, we did not directly survey students about their experiences with Drafter (although the course routinely collects a large amount of motivational data from students).
Instead, we focused on successfully deploying Drafter into the courses (ensuring that the library worked for our students), with the intention of exploring Drafter's motivational impact more deeply in the future.
Part of our success criteria was whether the students were able to keep to the scheduled milestones.
The majority of students were able to create a project that met our specifications, without any reports of serious difficulty or burden on the course staff.

Because things went so smoothly, the instructors agreed that the final project's requirements could be increased.
Students should be more strictly graded on their documentation, unit test quality and coverage, and other code quality aspects.
The technical requirements for the project should also be increased.
For instance, we assumed that the inclusion of the list in their state would motivate students to incorporate \texttt{for} loop constructs, but this was not the case.
Instead, in the future, we will require students to use these constructs and to have more complicated nested data (which aligns with the course's other material).
Generally, we think there is an open question about how to calibrate the difficulty of an open-ended final project for a library like Drafter, and believe this to be an interesting research question for future study.

Another issue we identified was that Drafter came relatively late into the curriculum, in the second half of the course.
At several points earlier in the course, we feel that there are natural integration points (possibly as early as week 2 of the semester, but certainly once we have taught basic dataclasses and lists in week 4).
Fundamentally, Drafter requires functions, so it would be difficult to teach Drafter at the very beginning of the semester without some hand-waving (although that could also be effective at foreshadowing to students what is ahead for them).
On the other side of the spectrum, there were assignments that we felt should come before the final project.
In particular, our course's curriculum has an existing lesson explaining the basics of accessing web data through the use of the \texttt{requests} library.
The instructors believe that moving this lesson before the final project would help students understand the context a little better.


There were a few issues when students went to deploy their sites, although the overall workflow was effective.
First, students generally did not write clear commit messages.
Second, they were instructed to modify a few key files (e.g., the \texttt{Readme.md} file), but some students ignored these instructions as well.
We have subsequently enhanced the deployment scripts to validate that the students have at least made some modifications to the appropriate file and avoided default commit messages, alerting them (and the graders) if this was not the case.
In fact, these scripts represent a tremendous opportunity to provide graders with more automated insight into students' code.
For instance, coverage reports are automatically generated, along with results from static analyses and typechecking tools like MyPy.
And most critically, deployment errors can be elegantly caught and shown, helping students diagnose and navigate the GitHub Actions deployment screens (which were a little obtuse for the folks who were struggling, who also happened to be the ones most likely to have errors).

We only had a few reports of technical difficulties with Drafter.
One recurring one was that students did not always understand where to put the \texttt{start\_server} call (supposed to be the very last call of the module).
Sometimes, students would place this call prematurely; since the call is blocking, any code (e.g., additional routes or unit tests) would not be executed.
We have not yet found an elegant solution for this problem, although Drafter could be modified to inspect their Python file and provide warnings if there is code left after a blocking \texttt{start\_server} call.

Some students had difficulty developing their \texttt{State} class and choosing appropriate fields.
Although some of this was the innate complexity of nested application state, some students were also stymied by inherent limitations of Dataclasses.
The biggest example of this was students who wanted to provide default values for list fields in their \texttt{State}, which requires the use of a specialized Python feature in the Dataclasses library (and was not taught as part of the curriculum).
Drafter can do more to inspect the design of students' \texttt{State} and offer more concrete advice, such as providing the extra code necessary to properly initialize mutable fields, warn about constant fields, and suggest proper naming convention for fields.


\section{Conclusion}

In conclusion, Drafter represents a significant step forward in making web development accessible and educationally valuable for novice computer science students.
By addressing the complexities and cognitive load typically associated with learning HTML, CSS, JavaScript, and various backend technologies, Drafter enables students to focus on core programming principles using Python.
Its design promotes good practices such as functional programming, decomposition, and unit testing while discouraging reliance on global mutable state.
Our initial deployment of Drafter in CS1 courses demonstrated that students could successfully use it to create meaningful projects, thereby bridging the gap between introductory programming education and real-world web development.
As we refine the curriculum and continue to enhance Drafter’s capabilities, we anticipate even greater integration and effectiveness in teaching fundamental computer science concepts through the engaging context of web development.

\bibliographystyle{ACM-Reference-Format}
\balance
\bibliography{sample-base}

\end{document}